\documentclass[manuscript]{aastex}

\begin{document}

\title{`Similar' coordinate systems and the Roche geometry. Application}
\shorttitle{`Similar' coordinate systems}
\shortauthors{Rodica Roman}

\author{Rodica Roman}
\affil{Astronomical Institute of Romanian Academy, Astronomical Observatory Cluj-Napoca,
               Str. Ciresilor 19, RO-400487 Cluj-Napoca, Romania }
\email{rdcroman@yahoo.com}

\begin{abstract}
A new equivalence relation, named relation of 'similarity' is defined and applied in the restricted three-body problem. Using this relation, a new class of trajectories (named 'similar' trajectories) are obtained; they have the theoretical role to give us new details in the restricted three-body problem.
The `similar' coordinate systems allow us in addition to obtain a unitary and an elegant demonstration of some analytical relations in the Roche geometry. As an example, some analytical relations published in Astrophysical Journal by Seidov in 2004 are demonstrated.
\end{abstract}

\keywords{Restricted problems: restricted problem of three bodies --- Stellar systems: binary stars }

\section{Introduction}
In the frame of the restricted three-body problem, the Roche geometry is a fundamental notion,
 and it is much studied, using for this aim different coordinate systems.

The usual way of treating the problem of motion of test particle and the Roche geometry in the 
gravitational field of a binary system consists in introducing a coordinate system (xyz), rotating 
jointly with the components. The x-axis passes through the centers of both components, and the y-axis 
is situated in the orbital plane; but there are more possibilities to locate the origin of the coordinate system. 
For example the origin can be located in the mass center of the binary system \citep{mo23}, or in the 
center of the more massive star \citep{ko78, ko89, ro88, sz67}. But there are some 
papers \citep{hu67,kr63}, where the location of the origin of the coordinate system is 
not precisely indicated. For example Huang (1967) wrote: "Thus, if we denote $1-\mu$ as the mass of one component,
$\mu $ will be the mass of the other. Let us further choose a rotating (x,y,z) system such that the origin is at 
the center of the $1-\mu$ component, the x-axis points always towards the $\mu $ component, and the $xy$ plane 
coincides with the orbital plane."

This way to locate the origin of the coordinate system is not an ambiguous one, but it offers
the opportunity of the question: What happens with the equations of motion of the test particle 
and with the geometry of the equipotential curves, in the restricted three-body problem, if the 
origin of the coordinate system is taken in the center of the less massive star.

The aim of this paper is to answer the question above, by introducing a new notion: the 'similar' 
coordinate systems.

In order to do so, a binary relation is created, which is denoted by the author as relation of 'similarity'.
Then, using this relation, the 'similar' coordinate systems, and the necessary 'similar' parameters and physical 
quantities are defined, obtaining 'similar' equations of motion, and 'similar' equipotential curves.

The conclusion of this study is that the use of 'similar' coordinate systems in the restricted 
three-body problem allows us to complete the traditional study of the Roche geometry with some 
new features.

\section{Relation of 'similarity'}
In this article we shall write 'similar' (not similar), because we intend to use this word as the name of a new mathematical relation, and not as an adjective.

\textit{Definition}: Two or more mathematical objects are in the
'similarity' relation in connection with a given definition
($\mathcal{D}$), if the objects are completely characterized by
$\mathcal{D}$.

For example:
\begin{enumerate}
    \item in algebra: The numbers $x_1=1$ and $x_2=-3$ are `similar' in connection with the definition: \textit{x is the solution of equation: $x^2+2x-3=0$, $x\in \mathbb{Z}$}.
    \item in geometry: If we consider the definition: \textit{P is the point situated into a given plane ($\pi$), at the distance $r=1$ from the given point A, $A\in(\pi)$}, then all the points of the circle having the center A and radius $r=1$ are 'similar'.
    \item in astrophysics: In the frame of the circular restricted three-body problem \textit{we define a comoving coordinate system situated in the orbital plane, having the origin in the center of one component of the binary system and the abscissa's axis pointing to the other component}. Then the coordinate systems $xS_1y$ and $x'S_2y'$ are `similar' (see Figure 1). (In many books  or articles of astronomy, this is the coordinate system's definition used to study the restricted three-body problem \citep{hu67}, but usualy only $xS_1y$ coordinate system is considered.
In 1963, Kruszewski wrote: "The center of coordinates is placed at the center of the (arbitrarily chosen) primary component" \citep{kr63}. This means that if the origin of the coordinate system is taken into the center of the secondary component, the results will be similar. In this article we try to find what this similarity implies.)
\end{enumerate}
 The study of the restricted three-body problem in 'similar' coordinate systems follows the classical algorithm, but some typical peculiarities appear. So, the use of the 'similar' coordinate systems impose the use of some physical and geometrical 'similar' quantities:
 \begin{description}
\item - 'similar' mass ratios $q$ and $q'$
\item - 'similar' distances $r_1$, $r_2$, $r_1'$, and $r_2'$
\item - 'similar' initial velocities $v_{0x}$, $v_{0y}$, $v_{0x}'$, and $v_{0y}'$
\item - 'similar' initial positions $x_0$, $y_0$, $x_0'$, and $y_0'$.
\end{description}
Of course, as in algebra the 'similar' solutions of a polynomial
equation are connected by the relations of Vi\`ete, the two
'similar' coordinate systems are connected by the equations of
coordinate transformations (see section 4).

It is easy to verify that the relation of 'similarity' is \textbf{reflexive}, \textbf{symmetric}, and \textbf{transitive}. That means that the relation of 'similarity' is an \textbf{equivalence relation}.

\textit{Remark}:
\begin{enumerate}
\item The role of the definition into the establishment of a 'similarity' is huge. So, if we consider the definition: \textit{x is the solution of equation: $x^2+2x-3=0$, $x\in \mathbb{N}$}, the numbers $x_1=1$ and $x_2=-3$ are not `similar'.
\item The solutions of a problem (including the repeated ones) are in relation of 'similarity', because these can be defined   as satisfying the same problem.
\end{enumerate}

\section{'Similar' coordinate systems, 'similar' parameters and physical quantities}

In the frame of the circular, restricted three-body problem \citep{sz67}, we will consider $S_1$ and $S_2$ the components of a
binary system, $m_1$ and $m_2$ their masses and $L_i$,
$i=\overline{1,5}$ the Lagrangian points. Due to the normalization
(see section 4), the distance between $S_1$ and $S_2$ is equal to 1. We will consider \textit{a rectangular
coordinate system, so that one component of the binary star has coordinates (0,0,0), the other
one has coordinates (1,0,0), and the angular Keplerian velocity $\vec{\omega}_k$ has components $(0, 0, \omega_k)$}. We observe immediately that there are
two coordinate systems: ($xS_1yz$) and ($x'S_2y'z'$), which can be
built. These are `similar' coordinate systems (see Figure 1).
\begin{figure}
  \includegraphics[height=.2\textheight]{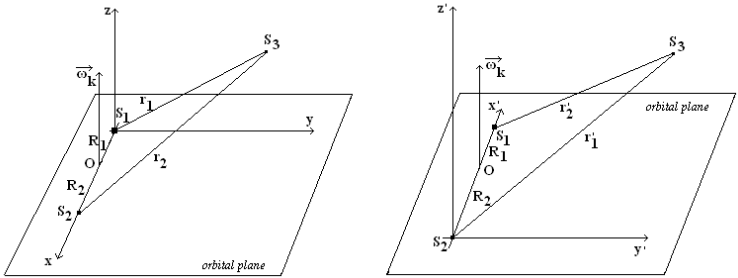}
  \caption{The 'similar' coordinate systems with origin in $S_1$ and $S_2$ .}
\end{figure}
As it is well-known \citep{ko78} p.327-328, \citep{ko89} p.15-16 the mass ratio $q$ is the main
parameter which describes the Roche geometry. If we denote
\textit{$q$ = (mass which isn't into the origin)/(mass which is
into the origin)}, we have $q=\frac{m_2}{m_1}$ and
$q'=\frac{m_1}{m_2}$ which are `similar' parameters. We denote
\textit{$r_1$ = distance of the infinitesimal mass} \citep{sz67} \textit{ to the origin of the coordinate system}, and \textit{$r_2$ =
distance of the infinitesimal mass to the star which is not into the origin}. Therefore $r_1$
with $r_1'$ and $r_2$ with $r_2'$ are `similar' distances (see Figure 1).

We have "'similar' velocities $v=\frac{dr_1}{dt}$ and $v'=\frac{dr_1'}{dt}$ and 'similar' accelerations $a=\frac{d^2r_1}{dt^2}$ and $a'=\frac{d^2r_1'}{dt^2}$.

\section{'Similar' equations of motion}

The forces which act on infinitesimal mass $S_3$ are $\vec{F}_{grav1}$, $\vec{F}_{grav2}$, $\vec{F}_{centrif}$, and $\vec{F}_{Coriolis}$ \citep{sz67} p.590). Their expressions in $(xS_1yz)$ and $(x'S_2y'z')$ coordinate systems are:
\begin{equation}
\vec{F}_{grav1}=-\frac{Gm_1m_3x}{r_1^3}\;\vec{i}-\frac{Gm_1m_3y}{r_1^3}\;\vec{j}-\frac{Gm_1m_3z}{r_1^3}\;\vec{k}
\end{equation}

\begin{equation}
\vec{F}_{grav2}=-\frac{Gm_2m_3[x-(R_1+R_2)]}{r_2^3}\;\vec{i}-\frac{Gm_2m_3y}{r_2^3}\;\vec{j}-\frac{Gm_2m_3z}{r_2^3}\;\vec{k}
\end{equation}

\begin{equation}
\vec{F}_{centrif}=m_3\,\omega_k^2\left[ (x-R_1)\vec{i}+y\vec{j} \right]
\end{equation}

\begin{equation}
\vec{F}_{Coriolis}=2m_3\,\omega_k\left( \frac{dy}{dt}\vec{i}-\frac{dx}{dt}\vec{j} \right)
\end{equation}
and respectively:
\begin{equation}
\vec{F'}_{grav1}=-\frac{Gm_2m_3x'}{r_1^{'3}}\;\vec{i}-\frac{Gm_2m_3y'}{r_1^{'3}}\;\vec{j}-\frac{Gm_2m_3z'}{r_1^{'3}}\;\vec{k}
\end{equation}

\begin{equation}
\vec{F'}_{grav2}=-\frac{Gm_1m_3[x'-(R_1+R_2)]}{r_2^{'3}}\;\vec{i}-\frac{Gm_1m_3y'}{r_2^{'3}}\;\vec{j}-\frac{Gm_1m_3z'}{r_2^{'3}}\;\vec{k}
\end{equation}

\begin{equation}
\vec{F'}_{centrif}=m_3\,\omega_k^2\left[ (x'-R_2)\vec{i}+y'\vec{j} \right]
\end{equation}

\begin{equation}
\vec{F'}_{Coriolis}=-2m_3\,\omega_k\left( \frac{dy'}{dt}\vec{i}-\frac{dx'}{dt}\vec{j} \right)
\end{equation}
where $G$ is the gravitational constant.

$\vec{F}_{Coriolis}$ and $\vec{F'}_{Coriolis}\,$ have opposite signs because of the orientation of vectors  $\vec{\omega}_k\times \vec{v}$ and $\vec{\omega'}_k\times \vec{v'}$ (see Figure 2).

\begin{figure}
  \includegraphics[height=.20\textheight]{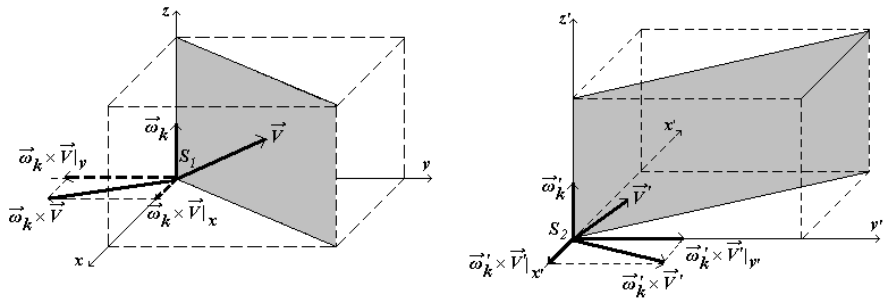}
  \caption{Explanation for the opposite signs of Coriolis force in the two 'similar' coordinate systems .}
\end{figure}

By consequence in the $(xS_1yz)$ coordinate system the vectorial equation of motion is:
\begin{equation}
m_3\:\vec{a}=\vec{F}_{grav1}+\vec{F}_{grav2}+\vec{F}_{centrif}+\vec{F}_{Coriolis}
\end{equation}
and in the $(x'S_2y'z')$ coordinate system:

\begin{equation}
m_3\:\vec{a'}=\vec{F'}_{grav1}+\vec{F'}_{grav2}+\vec{F'}_{centrif}+\vec{F'}_{Coriolis}
\end{equation}

We shall use a special unit system \citep{ko78} p.318:
we choose for the mass unit the sum of the masses of the components of the binary system, for the distance unit the distance between the centers of the components, and for the time unit the reciprocal of the angular Keplerian velocity.
In that case, the orbital period will be $P=2\pi$, and $G=1$.

Then $m_1=\frac{1}{1+q}$, $m_2=\frac{q}{1+q}$, $R_1=\frac{q}{1+q}$, $R_2=\frac{1}{1+q}$.

Using the same unit system, the scalar equations of motion in the $(xS_1yz)$ coordinate system become:
\begin{equation}
\frac{d^2x}{dt^2}-2\frac{dy}{dt}=x-\frac{q}{1+q}-\frac{x}{(1+q)r_1^3}-\frac{q(x-1)}{(1+q)r_2^3}
\end{equation}

\begin{equation}
\frac{d^2y}{dt^2}+2\frac{dx}{dt}=y-\frac{y}{(1+q)r_1^3}-\frac{q\:y}{(1+q)r_2^3}
\end{equation}

\begin{equation}
\frac{d^2z}{dt^2}=-\frac{z}{(1+q)r_1^3}-\frac{q\:z}{(1+q)r_2^3}\;,
\end{equation}
where $r_1=\sqrt{x^2+y^2+z^2}\,\;,\;\;r_2=\sqrt{(x-1)^2+y^2+z^2}\;.$

The equations of motion in the $(x'S_2y'z')$ coordinate system become:
\begin{equation}
\frac{d^2x'}{dt^2}+2\frac{dy'}{dt}=x'-\frac{q'}{1+q'}-\frac{x'}{(1+q')r_1^{'3}}-\frac{q'(x'-1)}{(1+q')r_2^{'3}}
\end{equation}

\begin{equation}
\frac{d^2y'}{dt^2}-2\frac{dx'}{dt}=y'-\frac{y'}{(1+q')r_1^{'3}}-\frac{q'\:y'}{(1+q')r_2^{'3}}
\end{equation}

\begin{equation}
\frac{d^2z'}{dt^2}=-\frac{z'}{(1+q')r_1^{'3}}-\frac{q'\:z'}{(1+q')r_2^{'3}}\;,
\end{equation}
where $r_1'=\sqrt{x'^2+y'^2+z'^2}\,\;,\;\;r_2'=\sqrt{(x'-1)'^2+y'^2+z'^2}\;.$

It can be easily verified that the equations of coordinate transformation are:
\begin{equation}
x'=1-x\;,\;\;y'=y\;,\;\;z'=z\,.
\end{equation}

\section{'Similar' initial conditions}
We have three differential equations of motion of second degree, therefore the initial conditions consist in three initial positions and three initial velocities. But in the following we shall consider the motion of the test particle in the orbital plane, so the equations of motion will be (11)-(12) and (14)-(15) respectively. By consequence we need two initial positions and two initial velocities.

\subsection{'Similar' initial positions}
Considering $P_0$ the point corresponding to the initial position, we denote $\overline{P_0}$ the projection of this point to the abscissa's axis.

Following the same idea as in section 3, we define {\bfseries the
initial abscissa} as \textit{a given number whose absolute value
represents the distance of $\overline{P_0}$ to the more massive
star, measured in a given sense of abscissa's axis}.

We define {\bfseries the initial ordinate} as \textit{a given number representing the distance $P_0\overline{P_0}$} (see Figure 3, where $S_1$ was considered the most massive star).

By consequence we have two 'similar' initial positions: $P_0(x_0,y_0)$ and $P_0'(x_0',y_0')$, where
\begin{equation}
x_0'=1+x_0\;,\;\;\;y_0'=y_0\;.
\end{equation}

\begin{figure}
  \includegraphics[height=.30\textheight]{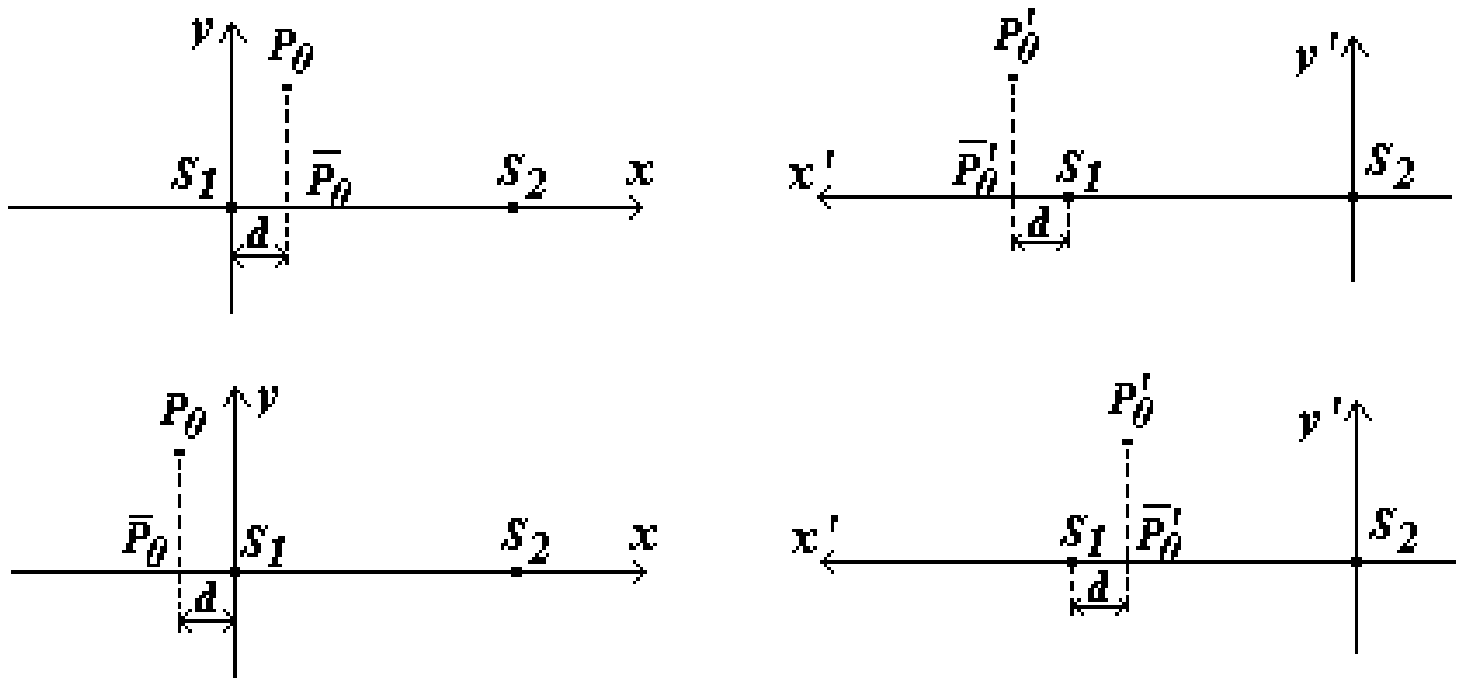}
  \caption{'Similar' initial abscissa.}
\end{figure}

\subsection{'Similar' initial velocities}
We define the initial velocity using the following three conditions:
\textit{\begin{enumerate}
    \item The components of initial velocity have given absolute value.
    \item The abscissa's axis, the initial velocity vector, and the Keplerian angular velocity vector form a trihedron with a given orientation (positive or negative).
    \item The angle formed by the initial velocity vector and the positive abscissa's semi-axis has a
     given type (acute or obtuse).
\end{enumerate}}

Then, for $V_0(v_{0x},v_{0y})$ and  $V_0'(v_{0x}',v_{0y}')$ we have:
\begin{equation}
v_{0x}'=-v_{0x}\;,\;\;\;v_{0y}'=-v_{0y}\;.
\end{equation}

\textit{Remark}:
All three conditions are necessary. If only (i) is considered, we
have four possibilities for $V_0$ (see Figure 4 (a)). If the conditions (i) and (ii) are considered, there are two possibilities for $V_0$ (see Figure 4 (b), where a positive orientation of the trihedron is taken). If all the three conditions are considered, there is only one possibility (see Figure 4 (c) for an acute angle formed by the initial velocity vector and the positive abscissa's semi-axis).
\begin{figure}
  \includegraphics[height=.50\textheight]{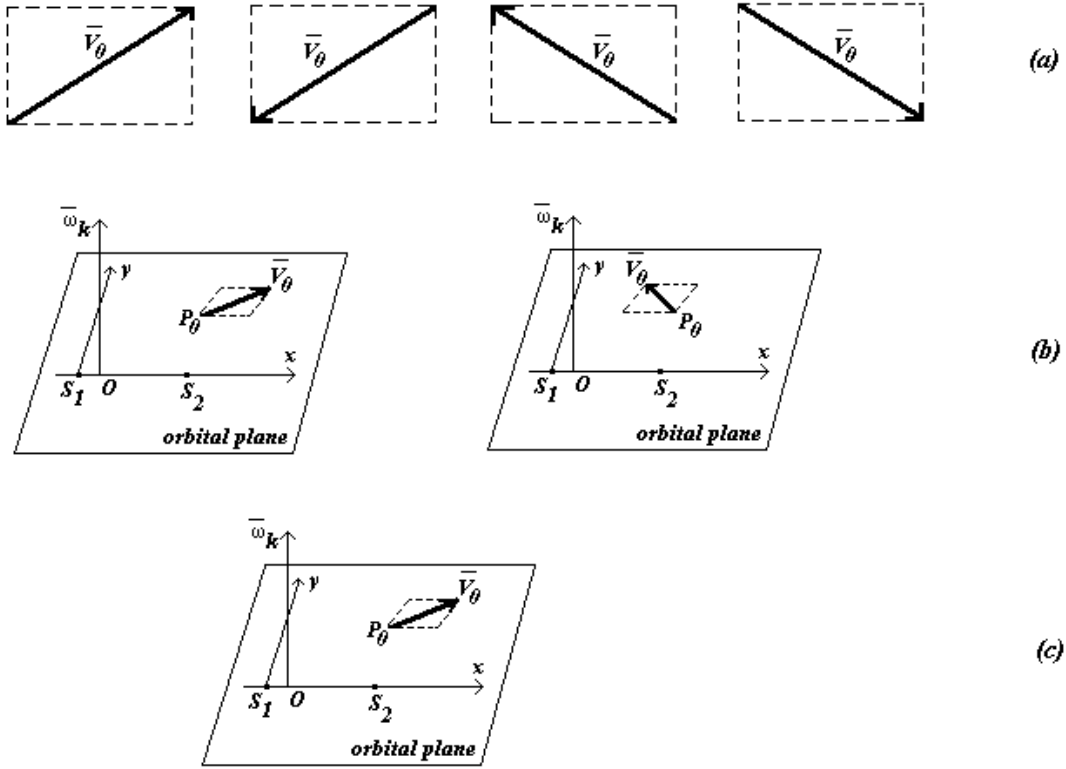}
   \caption{'Similar' initial velocity .}
\end{figure}

\section{'Similar' trajectories}

As a numerical application in Figure 5, there are given two 'similar' trajectories for the binary system Earth-Moon ( $q=0.0121$ and $q'=82.45$). The initial conditions are: $x_0=-0.2\;,\;\;y_0=-0.5\;,\;\;v_{0x}=-0.8\;,\;\;v_{0y}=-0.6\;,$ respectively $x_0'=0.8\;,\;\;y_0'=-0.5\;,\;\;v_{0x}'=0.8\;,\;\;v_{0y}'=0.6\;$. The time of integration is one Keplerian period.
\begin{figure}
  \includegraphics[height=.25\textheight]{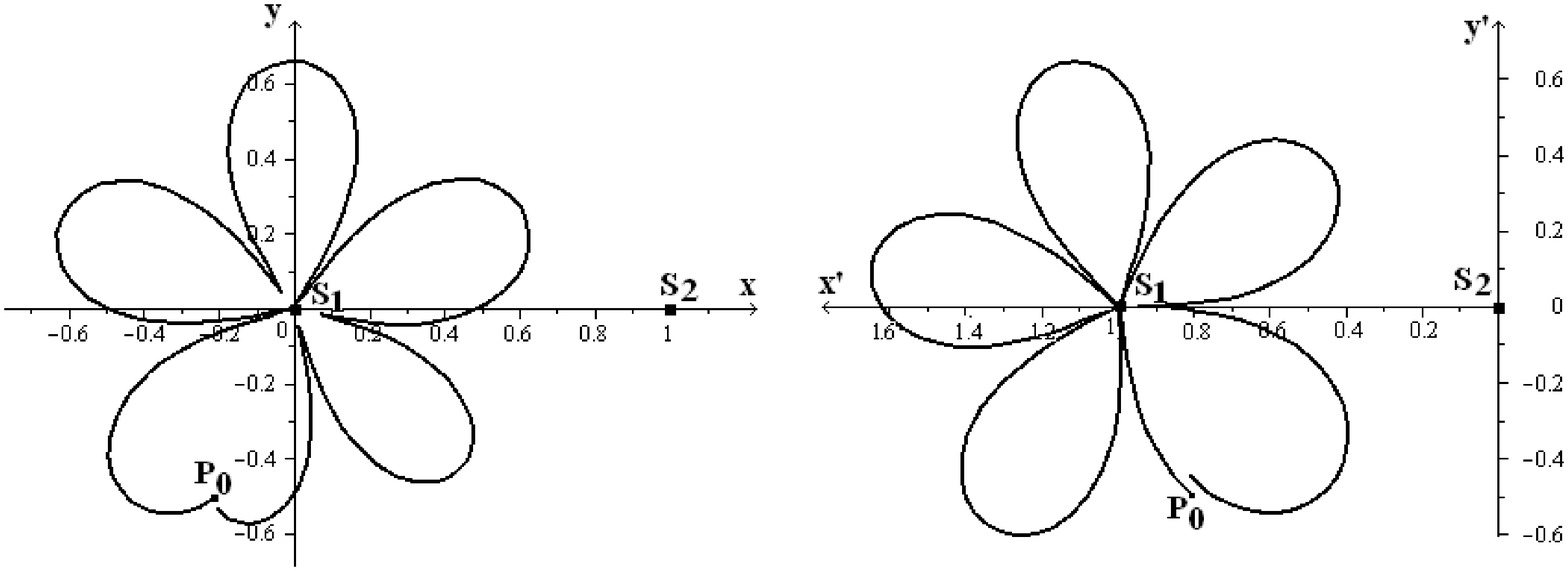}
   \caption{'Similar' trajectories .}
\end{figure}
In Figure 6, the 'similar' trajectories for the
same binary system, with initial conditions:
$x_0=0.6\;,\;\;y_0=0.4\;,\;\;v_{0x}=0.5\;,\;\;v_{0y}=0\;,$
respectively
$x_0'=1.6\;,\;\;y_0'=0.4\;,\;\;v_{0x}'=-0.5\;,\;\;v_{0y}'=0\;$ are presented.
The time of integration is one Keplerian period. In Figure 7,
there are given the 'similar' trajectories for the same conditions
as in Figure 6, but the time of integration is 200 Keplerian
periods.
\begin{figure}
  \includegraphics[height=.3\textheight]{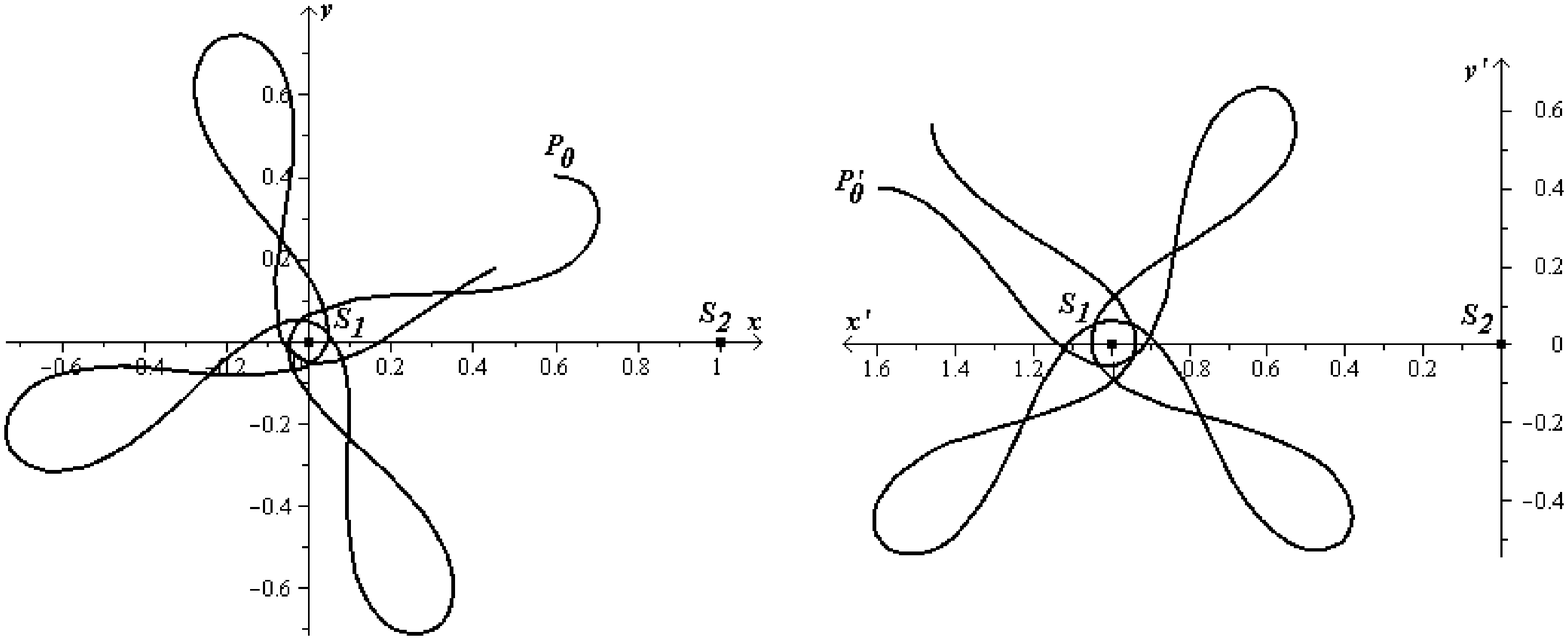}
   \caption{Others 'similar' trajectories .}
\end{figure}

\begin{figure}
  \includegraphics[height=.3\textheight]{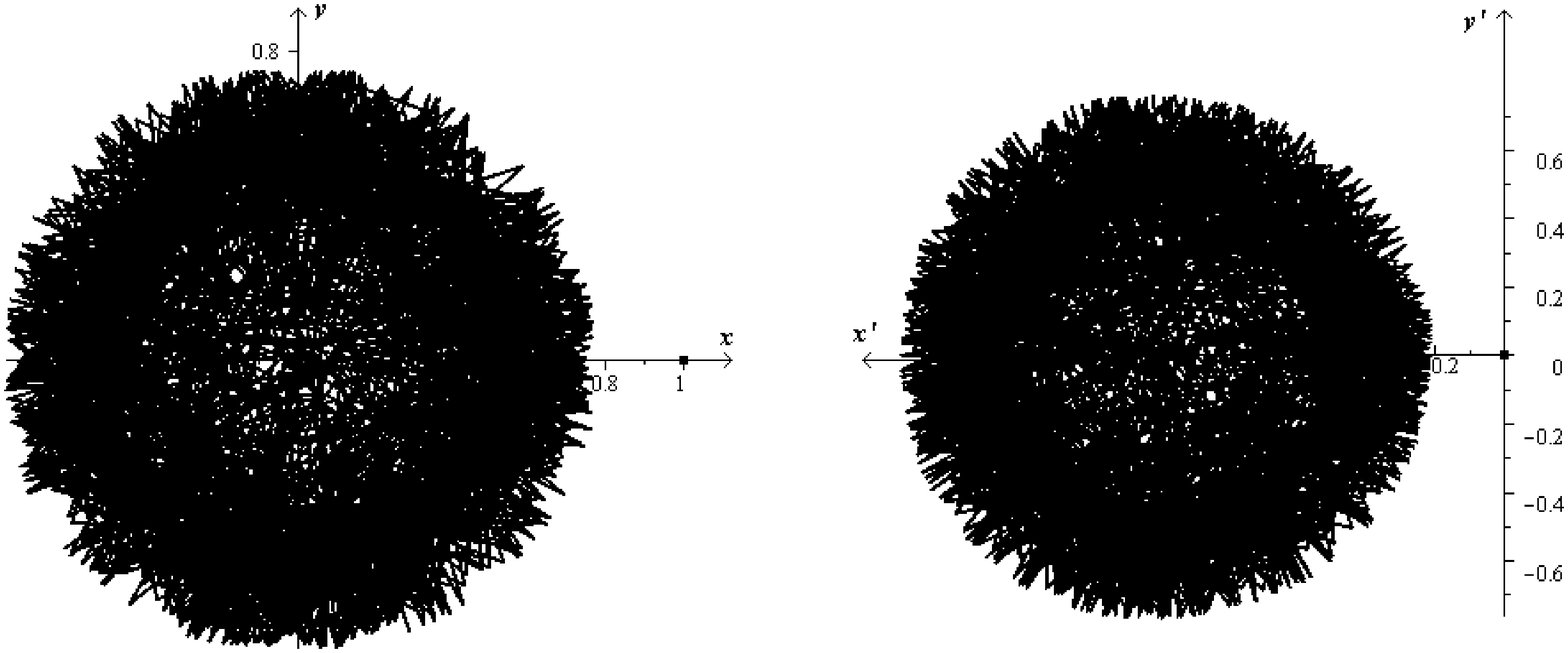}
   \caption{ 'Similar' trajectories for 200 Keplerian period time of integration.}
\end{figure}

\section{The use of 'similar' coordinate systems to demonstrate some analytical relations in the Roche geometry}

We shall show now the consequence of 'similar' coordinate system use on the equipotential curves (see Figure 8), and on the equilibrium points' positions, in the frame of the Roche geometry.

From equations of motion (11)-(12) we obtain the equation of potential function:
$$\Psi (x,y)=\frac{1}{2}\left[ \left( x-\frac{q}{1+q} \right)^2+y^2+\frac{2}{(1+q)r_1}+\frac{2q}{(1+q)r_2}  \right] \;.$$
(We prefer to use this form, because in the second part of this
paragraph we shall demonstrate some analytical relations of \cite{se04}, who denoted the expression in square brackets with $\psi
(x,y)$, (see equation 21).) Similarly, from equations
(14)-(15), the corresponding potential function is:
$$\Psi' (x',y')=\frac{1}{2}\left[ \left( x'-\frac{q'}{1+q'} \right)^2+y'^2+\frac{2}{(1+q')r_1'}+\frac{2q'}{(1+q')r_2'}  \right] \;.$$

In Figure 8 are represented the equipotential curves for a binary system characterized by the mass ratio $q=0.2$ (and by consequence $q'=5$). Let us consider $L_i(x_i,0)$, $i=\overline{1,5}$ the Lagrangian points. For the first Lagrangian point we obtained $x_1=0.65856$ and $x_1'=0.34144$. The Jacobian constant ( \citep{sz67} p.16 is $C_1=C_1'=3.74897$. For the second Lagrangian point, $x_2=1.43808$ and $x_2'=-0.43808$ and $C_2=C_2'=3.53634$. For the third Lagrangian point, $x_3=-0.90250$ and $x_3'=1.90250$ and $C_3=C_3'=3.16504$. The points $L_4$ and $L_5$ form equilateral triangles with $S_1$ and $S_2$. The numerical results obtained above are normal, because $x'=1-x$ (see equation(17)). In what concerne $C_i=C_i'\;,\;\;i=\overline{1,3}$, the equalities are normal if we think at the physical meaning of the Jacobian constant. In \cite{se04}, there are obtained some analytical relations in the Roche geometry and one of them can be considered as an analytical demonstration of the relation $C_i=C_i'\;,\;\;i=\overline{1,3}$. He obtained a very important correlation between
the potential ($\psi$) and the mass ratio ($q$) on one side, and
the Lagrangian points $L_1(x_1,0)$ and $L_2(x_2,0)$ on the other
side, in the frame of the classical Roche model. The relations
obtained by Seidov are:
\begin{eqnarray}
& & \psi_1(q,x_1)=\psi_1\left(1/q, 1-x_1\right)\,,{}
\nonumber \\& & {} q(x_i)\cdot q(1-x_i)=1\,, \;\;i=\overline{1,2}.
\end{eqnarray}
The first relation of (20) correspond to $C_i=C_i'\;,\;\;i=\overline{1,2}$ .

This paragraph will show an elegant and unitary proof for formulae (20), using the `similar' coordinate systems. Using $q$, $q'$, $r_1$, $r_1'$, $r_2$, $r_2'$ we will obtain these
relations for $i=\overline{1,3}$ and we will expand the
expressions concerning the potential for $i=\overline{4,5}$. So,
the analytical relations of Seidov will be analyzed for all
Lagrangian points, in a very easy manner.

\begin{figure}[!t]
  \includegraphics[height=.30\textheight]{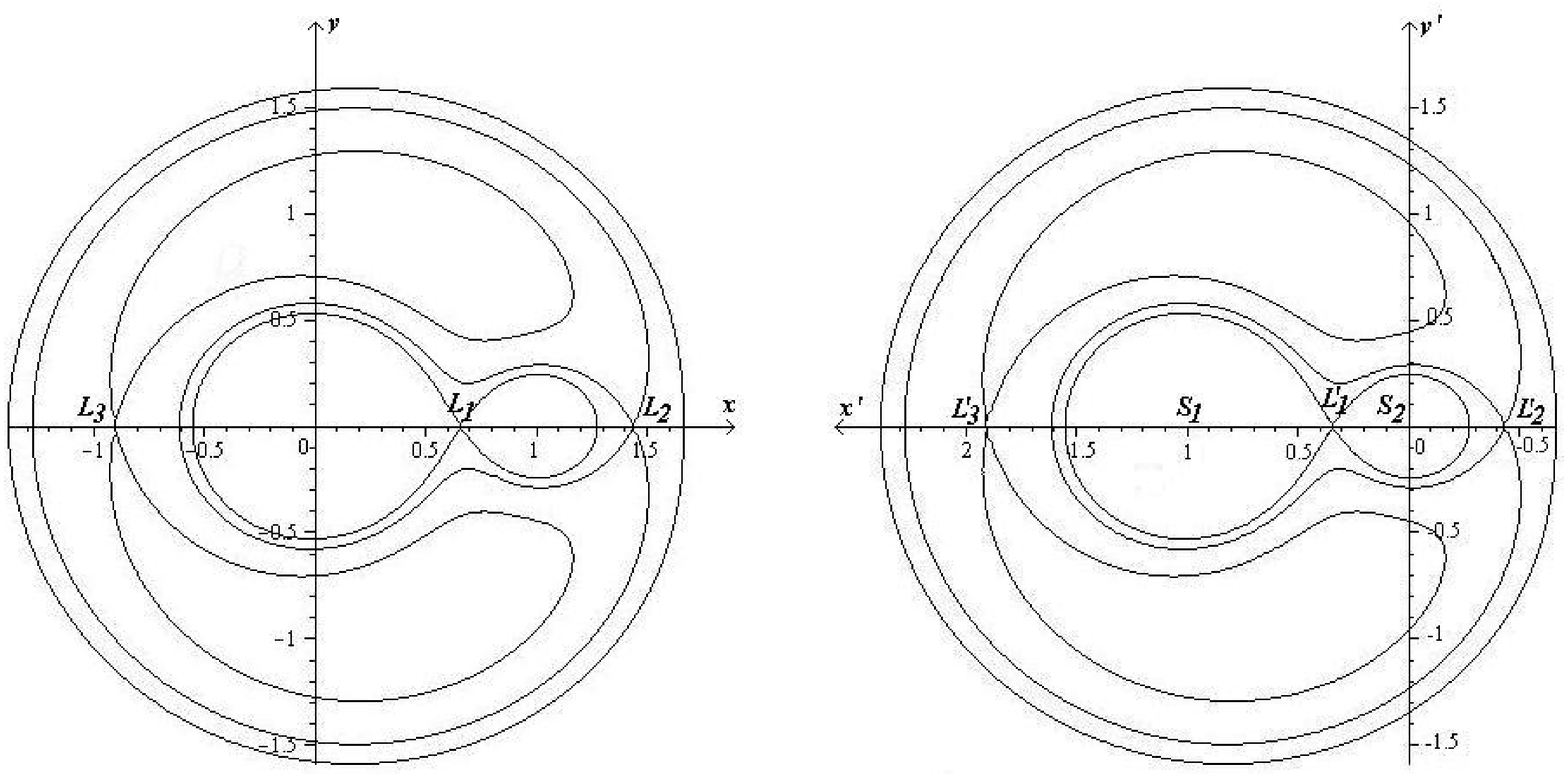}
  \caption{The `similar' equipotential curves for the Roche model}
 \end{figure}

\begin{figure}[!t]
  \includegraphics[height=.50\textheight]{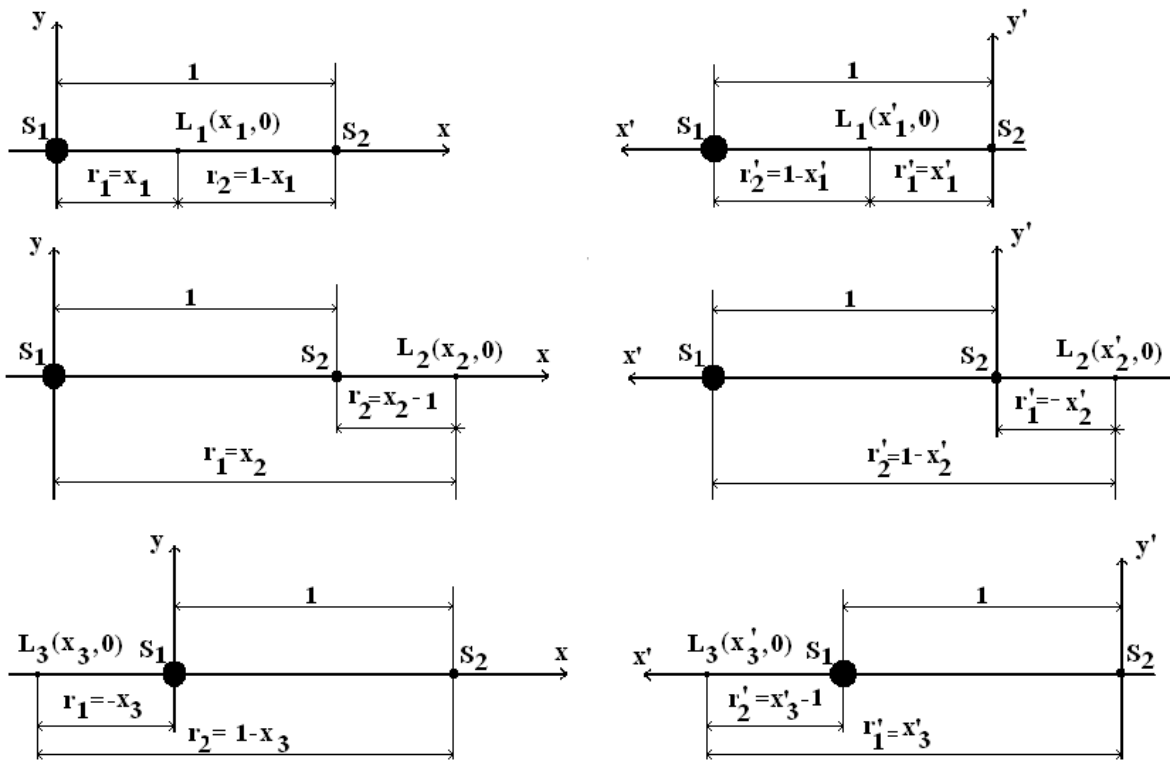}
  \caption{The `similar' coordinate systems and the Lagrangian points $L_1$, $L_2$, and $L_3$.}
 \end{figure}

The equation of the potential given in \cite{se04} in the $(xS_1y)$ system is:
\begin{eqnarray}
\psi(x,y)&=&\left(x-\frac{q}{1+q}\right)^2+y^2+{}
\nonumber\\& &{}\frac{2}{(1+q)r_1}+\frac{2q}{(1+q)r_2}
\end{eqnarray}
and if we use 'similar' coordinate systems,
the equation of the potential in $(x'S_2y')$ system becomes:
\begin{eqnarray}
\psi'(x',y')&=&\left(x'-\frac{q'}{1+q'}\right)^2+y'^2{}
\nonumber\\& &{}+\frac{2}{(1+q')r_1'}+\frac{2q'}{(1+q')r_2'}\;.
\end{eqnarray}
where $$r_1=\sqrt{x^2+y^2}\,,\;\; r_2=\sqrt{(1-x)^2+y^2}\,,$$
 $$r_1'=\sqrt{x'^2+y'^2}\,, \;\;r_2'=\sqrt{(1-x')^2+y'^2}\,.$$

$L_i(x_i,0)$ being equilibrium points, we have
$\partial\psi/\partial x_i=0$ and

$\partial\psi'/\partial x_i'=0$, $i=\overline{1,3}$ (see Figure 9). From these relations we obtain:
\begin{equation}
q(x_1)=\frac{(1-x_1)^3(1+x_1+x_1^2)}{x_1^3(3-3x_1+x_1^2)}\;,
\end{equation}
\begin{equation}
q'(x_1')=\frac{(1-x_1')^3(1+x_1'+x_1'^2)}{x_1'^3(3-3x_1'+x_1'^2)}\;,
\end{equation}
\begin{equation}
q(x_2)=\frac{(x_2-1)^3(1+x_2+x_2^2)}{x_2^2(2-x_2)(1-x_2+x_2^2)}\;, 
\end{equation}
\begin{equation}
q'(x_2')=-\frac{(1-x_2')^2(x_2'^3+1)}{x_2'^3(3-3x_2'+x_2'^2)}\;,
\end{equation}
\begin{equation}
q(x_3)=-\frac{(1-x_3)^2(x_3^3+1)}{x_3^3(3-3x_3+x_3^2)}\;,
\end{equation}
\begin{equation}
q'(x_3')=\frac{(1-x_3')^2(x_3'^3-1)}{x_3'^2(2-x_3')(1-x_3'+x_3'^2)}\;.
\end{equation}

From (17) one can observe that  $x_i'=1-x_i\,, \;i=\overline{1,3}$, and therefore $q(x_i)=\frac{1}{q'(1-x_i)}$, and $q(x_i)\cdot q'(1-x_i)=1\;, i=\overline{1,3}$. These relations can be compared with the relations (20) given in \cite{se04}, for $i=\overline{1,2}$.

For $i=\overline{4,5}$ we have $L_4(\frac{1}{2},\frac{\sqrt{3}}{2})$,  $L_5(\frac{1}{2},-\frac{\sqrt{3}}{2})$, \citep{ro88, mu05}, (the coordinates of $L_4$ and $L_5$ are independent of $q$), therefore $q$ is not a  function of $x_i$, $i=\overline{4,5}$. Here $r_1=1=r_2$.

From (20) and (21), the potential corresponding to the Lagrangian points are:
\begin{eqnarray}
\psi(q,x_1)&=&\left(x_1-\frac{q}{1+q}\right)^2+{}
\nonumber\\& &{}\frac{2}{(1+q)x_1}+\frac{2q}{(1+q)(1-x_1)}\;,
\end{eqnarray}
\begin{eqnarray}
\psi'(q',x_1')&=&\left(x_1'-\frac{q'}{1+q'}\right)^2+{}
\nonumber\\& &{}\frac{2}{(1+q')x_1'}+\frac{2q'}{(1+q')(1-x_1')}\;,
\end{eqnarray}
\begin{eqnarray}
\psi(q,x_2)&=&\left(x_2-\frac{q}{1+q}\right)^2+{}
\nonumber\\& &{}\frac{2}{(1+q)x_2}-\frac{2q}{(1+q)(1-x_2)}\;,
\end{eqnarray}
\begin{eqnarray}
\psi'(q',x_2')&=&\left(x_2'-\frac{q'}{1+q'}\right)^2-{}
\nonumber\\& &{}\frac{2}{(1+q')x_2'}+\frac{2q'}{(1+q')(1-x_2')}\;,
\end{eqnarray}
\begin{eqnarray}
\psi(q,x_3)&=&\left(x_3-\frac{q}{1+q}\right)^2-{}
\nonumber\\& &{}\frac{2}{(1+q)x_3}+\frac{2q}{(1+q)(1-x_3)}\;,
\end{eqnarray}
\begin{eqnarray}
\psi'(q',x_3')&=&\left(x_3'-\frac{q'}{1+q'}\right)^2+{}
\nonumber\\& &{}\frac{2}{(1+q')x_3'}-\frac{2q'}{(1+q')(1-x_3')}\;.
\end{eqnarray}
Using the equations: $x_i'=1-x_i$, and  $q'=1/q$ we obtain:
$$\psi (q,x_i)=\psi '(1/q, 1-x_i)\;,\qquad i=\overline{1,3}\;.$$
These relations can be compared with relations (20) given in \cite{se04} for $i=\overline{1,2}$.

For $L_4(\frac{1}{2},\frac{\sqrt{3}}{2})$, using the equations (20) and (21) we obtain
$$\psi(q)=\frac{3+5q+3q^2}{(1+q)^2}\;,\qquad \psi'(q')=\frac{3+5q'+3q'^2}{(1+q')^2}$$
and because $q'=1/q$, for $L_4$ we obtain $\psi(q)=\psi'(1/q)$.

For $L_5(\frac{1}{2},-\frac{\sqrt{3}}{2})$ the same equation is obtained.

From equations (23)-(28) and (29)-(34) we obtain analytical
formulae for the potential as a function of Lagrangian point
positions:
$$\psi (x_1)=-\frac{4x_1^8-16x_1^7+14x_1^6+14x_1^5-41x_1^4+40x_1^3-27x_1^2+12x_1-3}{(x_1^4-2x_1^3-x_1^2+2x_1-1)^2}$$
$$\psi (x_2)=\frac{4x_2^7-14x_2^6+18x_2^5+9x_2^4-36x_2^3+27x_2^2-4x_2-1}{(x_2^4-2x_2^3-x_2^2+2x_2-1)^2}$$
$$\psi (x_3)=-\frac{4x_3^7-14x_3^6+18x_3^5-29x_3^4+40x_3^3-27x_3^2+12x_3-3}{(x_3^4-2x_3^3+x_3^2-2x_3+1)^2}$$
$$\psi'(x_1')=-\frac{4x_1'\,^8-16x_1'\,^7+14x_1'\,^6+14x_1'\,^5-41x_1'\,^4+40x_1'\,^3-27x_1'\,^2+12x_1'\,-3}{(x_1'\,^4-2x_1'\,^3-x_1'\,^2+2x_1'\,-1)^2}$$
$$\psi '(x_2')=-\frac{4x_2'\,^7-14x_2'\,^6+18x_2'\,^5-29x_2'\,^4+40x_2'\,^3-27x_2'\,^2+12x_2'\,-3}{(x_2'\,^4-2x_2'\,^3+x_2'\,^2-2x_2'\,+1)^2}$$
$$\psi'(x_3')=\frac{4x_3'\,7-14x_3'\,^6+18x_3'\,^5+9x_3'\,^4-36x_3'\,^3+27x_3'\,^2-4x_3'\,-1}{x_3'\,^4-2x_3'\,^3+x_3'\,^2+2x_3'\,-1}$$
The equation for $\psi(x_2)$ is obtained also by Seidov (see equation (9) in \cite{se04}).
\section{Conclusion}
The 'similarity' relation defined in section 2 belongs to equivalence relations' family. For the time being it has only a theoretical value, completing classical method of study of the restricted three-body problem (see sections 4, 5, 6), and allowing for a more elegant demonstration of some analytical relations from the geometry of the Roche model (section 7). The use of 'similar' coordinate systems imposes the typical definitions of mass ratio, of distances from the test particle to the components of the binary system, and of initial conditions necessary to integrate the differential equations of motion. The 'similar' trajectories are not like-wise, but have a similar topology.

The use of `similar' coordinate systems helped us to create an elegant and easy proof for the analytical relations obtained by \cite{se04}. To close the circle, we have completed these relations by analyzing the problem of mass ratio and potential, as function of Lagrangian point positions for all five Lagrangian points. So, the use of 'similar' coordinate systems in the restricted three-body problem enable us to complete the study of the Roche geometry with some new elements.

\acknowledgments
The author is very grateful to the anonymous reviewer for the attentive read of the manuscript and his valuable suggestions.


\begin{thebibliography}{}
\bibitem[Huang (1967)]{hu67} Huang, S. Sh., in "Modern Astrophysics. A memorial to Otto Struve", ed. M. Hack, Gauthier-Villars, Paris, Gordon and Breach, New York, p.211 (1967)
\bibitem[Kitamura(1970)]{ki70} Kitamura, M., Ap \& SS, 7, 272 (1970)
\bibitem[Kopal(1989)]{ko89} Kopal, Z., The Roche Problem, Dordrecht: Kluwer (1989)
\bibitem[Kopal(1978)]{ko78} Kopal, Z., Dynamics of Close Binary Systems, Dordrecht: Reidel (1978)
\bibitem[Kruszewski (1963)]{kr63} Kruszewski, A., Acta Astron., 13, 106 (1963)
\bibitem[Moulton (1923)]{mo23}Moulton, F. R., An introduction to celestial mechanics, Second Edition, The Macmillan Company, New York (1923)
\bibitem[Murray(2005)]{mu05} Murray, C. D., \& Dermott, S. F., Solar System Dynamics, Cambridge:  Cambridge University Press (2005)
\bibitem[Roy(1988)]{ro88} Roy, A. E., Orbital Motion, Bristol: Adam Hilger (1988)
\bibitem[Seidov(2004)]{se04} Seidov, Z. F., \apj, 603, 283 (2004)
\bibitem[Szebehely(1967)]{sz67} Szebehely, V., Theory of Orbits, New York: Academic Press (1967)
\end{thebibliography}
\end{document}